\documentclass[12pt]{article}
\usepackage{times}
\usepackage{amsfonts}
\topmargin - 0.5 cm 
\begin{document}
\baselineskip = 0.60 true cm

\thispagestyle{empty}

\vspace{1.8cm} 
{\bf Phase operators, temporally 
stable phase states, 

mutually unbiased bases and exactly 
solvable quantum systems}

\vspace{1.5cm} 
{\bf M Daoud$^{1,2,3,4}$ and  M R Kibler$^{1,2,3}$}

\vspace{0.8cm}


$^1$ Universit\'e Lyon, F-69622, Lyon, France

$^2$ Universit\'e Claude Bernard Lyon-1, Villeurbanne, France

$^3$ CNRS/IN2P3, Institut de Physique Nucl\'eaire de Lyon, France

$^4$ Facult\'e des Sciences, Agadir, Morocco


\vspace{0.5cm}
E-mail : m\_daoud@hotmail.com

E-mail : m.kibler@ipnl.in2p3.fr

\vspace{3cm} 
{\bf Abstract}

\baselineskip=18pt
\medskip

We introduce a one-parameter generalized oscillator algebra ${\cal A}_{\kappa}$ 
(that covers the case of the harmonic oscillator algebra) and discuss its finite- 
and infinite-dimensional representations according to the sign of the parameter 
$\kappa$. We define an (Hamiltonian) operator associated with ${\cal A}_{\kappa}$ 
and examine the degeneracies of its spectrum. For the finite (when $\kappa < 0$) 
and the infinite (when $\kappa \geq 0$) representations of ${\cal A}_{\kappa}$, 
we construct the associated phase operators and build temporally stable 
phase states as eigenstates of the phase operators. To overcome the difficulties 
related to the phase operator in the infinite-dimensional case and to avoid the 
degeneracy problem for the finite-dimensional case, we introduce a truncation 
procedure which generalizes the one used by Pegg and Barnett for the harmonic 
oscillator. This yields a truncated generalized oscillator algebra 
${\cal A}_{\kappa, s}$, where $s$ denotes the truncation order. We construct two types 
of temporally stable states for ${\cal A}_{\kappa, s}$ (as eigenstates 
of a phase operator and as eigenstates of a polynomial in the generators of 
${\cal A}_{\kappa, s}$). Two applications are considered in this article. The 
first concerns physical realizations of ${\cal A}_{\kappa}$ and 
${\cal A}_{\kappa, s}$ in the context of one-dimensional quantum systems with 
finite (Morse system) or infinite (P\" oschl-Teller system) discrete spectra. The 
second deals with mutually unbiased bases used in quantum information.

\newpage
\section{Introduction}

It is well known that the usual model for the quantized single modes of the electromagnetic field is 
the harmonic oscillator with an infinity of states. The infinite-dimensional character of the 
representation space of the corresponding oscillator algebra constitutes a drawback to define a phase 
operator in a consistent way \cite{Louisell63}-\cite{Susskind68}. In order to get rid of this difficulty, 
Pegg and Barnett suggested to truncate to some finite (but arbitrarily large) order the infinite-dimensional 
representation space of the oscillator algebra \cite{Pegg}. Their approach also provided a valid way 
for calculating the so-called phase states (the eigenvectors of the phase operator). In the same vein, 
Vourdas proposed a definition of a phase operator for $su(2)$ and calculated its eigenstates without a truncation 
procedure since $su(2)$ admits finite-dimensional unitary irreducible representations \cite{Vourdas90}. He also 
constructed a phase operator and its eigenstates for $su(1,1)$, without a truncation procedure although $su(1,1)$ 
admits infinite-dimensional unitary irreducible representations \cite{Vourdas90}. 

The main aim of the present work is to develop a method to build unitary phase operators\footnote{We deal 
here with {\it unitary} rather than {\it Hermitian} phase operators. The two kinds of operators are related 
via an exponentiation trick.} and temporally stable phase states for some exactly solvable quantum systems. Various 
algebraic structures were used to construct (temporaly stable or not) coherent states in connection with some quantum 
systems \cite{BarutGi}-\cite{Kinani}. The construction of temporally stable 
phase states to be developed in this work is based on a generalized oscillator algebra which takes its 
root in \cite{DaoKib12, DaoKibPLAJMP}. This algebra was introduced to construct isospectral 
shape invariant potentials in the framework of fractional supersymmetry. 

A second facet of this work is to show that the obtained temporally stable phase states can be used to generate mutually 
unbiased bases (MUBs). Such bases are of considerable interest in quantum information and were recently investigated from 
an angular momentum approach \cite{KibKPAlbKib, KibJPhysA0809}. It is not the purpose of this paper to deal with unsolved 
problems concerning MUBs but to give a way to construct MUBs from temporally stable states associated with some exactly 
solvable systems. 

The paper is organized as follows. Section 2 is devoted to the generalized oscillator algebra ${\cal A}_{\kappa}$. Temporally 
stable phase states associated with ${\cal A}_{\kappa}$ are studied in section 3. Section 4 deals with the truncated oscillator 
algebra ${\cal A}_{\kappa, s}$ and the correponding phase states. As a first application, the derivation of MUBs 
from phase states is developed in section 5. A second application is made in section 6 to some exactly solvable quantum systems. 

The notations are standard. Let us simply mention that: $\delta_{a , b}$ stands for the Kronecker symbol of $a$ and $b$, $I$ 
for the identity operator, $A^{\dagger}$ for the adjoint of the operator $A$, and $[A , B]$ and $\{ A, B \}$ for respectively 
the commutator and the anticommutator of the operators $A$ and $B$. We use a notation of type $\vert \psi \rangle$ for a vector 
in an Hilbert space and we denote $\langle \phi \vert \psi \rangle$ and $\vert \phi \rangle \langle \psi \vert$ respectively 
the inner and outer products of the vectors $\vert \psi \rangle$ and $\vert \phi \rangle$. 

\section{Generalized oscillator algebra}

\subsection{The algebra ${\cal A}_{\kappa}$}
Let ${\cal A}_{\kappa}$ be the algebra spanned by the three linear operators $a^-$, $a^+$ and $N$ 
satisfying the following relations
      \begin{eqnarray}
[a^- , a^+] = I + 2 \kappa N       \qquad 
[N , a^{\pm}] = \pm a^{\pm}        \qquad 
\left( a^- \right)^{\dagger} = a^+ \qquad
N^{\dagger} = N,                
      \label{thealgebra}
      \end{eqnarray}
where $\kappa$ is a real parameter. Note that, for $\kappa = 0$, the algebra ${\cal A}_{0}$ is nothing 
but the usual harmonic oscillator algebra. The operators $a^-$, $a^+$ and $N$ in (\ref{thealgebra})
generalize the annihilation, creation and number operators used for the harmonic oscillator. 
Therefore, the algebra ${\cal A}_{\kappa}$ shall be called generalized oscillator algebra. This algebra 
turns out to be a particular case of the generalized Weyl-Heisenberg algebra $W_k$ introduced in 
\cite{DaoKib12, DaoKibPLAJMP} and not to be confused with the Lie algebra of the Heisenberg-Weyl group $HW(\mathbb{R})$ 
used in quantum information \cite{KibJPhysA0809}. In fact, ${\cal A}_{\kappa}$ is identical to $W_k$ with
      \begin{eqnarray}
k = 1 \qquad f_0(N) = aN + b \qquad \frac{1}{\sqrt{b}} X_{\pm} = a^{\pm} \qquad \kappa = \frac{1}{2} \frac{a}{b},
      \label{lien avec Wk}
      \end{eqnarray}
where the operators $f_0(N)$ and $X_{\pm}$, and the parameters $k$, $a$ and $b$ are defined in \cite{DaoKibPLAJMP}. It
should be noted that the $C_{\lambda}$-extended oscillator algebra worked out in \cite{MD33LesQuene} is a particular case 
of $W_k$ (for $\lambda = k$). 

\subsection{Hilbertian representation of ${\cal A}_{\kappa}$}
We denote by ${\cal F}_{\kappa}$ the finite- or infinite-dimensional Hilbert space on which the 
operators $a^-$, $a^+$ and $N$ are defined. Let
\begin{eqnarray}
\{ \vert n \rangle : n = 0, 1, \ldots, d({\kappa}) \}
      \end{eqnarray}
(with $d({\kappa})$ finite or infinite) be an orthonormal basis, with respect to the inner product 
$\langle n \vert n' \rangle = \delta_{n,n'}$, of the space ${\cal F}_{\kappa}$. It is easy to check 
that the actions 
	  \begin{eqnarray} 
& & a^+ \vert n \rangle = \sqrt{F(n+1)} e^{{-i [F(n+1)- F(n)]  \varphi }} \vert n+1 \rangle, \nonumber \\ 
& & a^- \vert n \rangle = \sqrt{F(n)}   e^{{ i [F(n) - F(n-1)] \varphi }} \vert n-1 \rangle, \label{action sur les n} \\ 
& & a^- \vert 0 \rangle = 0 \qquad
    N   \vert n \rangle = n \vert n \rangle
    \nonumber 
	  \end{eqnarray} 
provide an Hilbertian representation of the algebra ${\cal A}_{\kappa}$ defined by (\ref{thealgebra}). In 
equation (\ref{action sur les n}), the real parameter $\varphi$ is arbitrary and the positively-valued function 
$F : \mathbb{N} \to \mathbb{R}_+$ satisfies the recurrence relation
	  \begin{eqnarray} 
F(n+1) - F(n) = 1 + 2 \kappa n \qquad F(0) = 0.
    \label{recurrence} 
	  \end{eqnarray} 
The iteration of (\ref{recurrence}) yields
	  \begin{eqnarray} 
F(n) = n [1 + \kappa (n - 1)], 
    \label{eigenvalue} 
	  \end{eqnarray} 
which is linear in $n$ only for $\kappa = 0$. Since $F(n) \in \mathbb{R}_+$, we must have the following condition 
	  \begin{eqnarray} 
1 + \kappa (n - 1) > 0  
    \label{condition} 
	  \end{eqnarray} 
for $n > 0$. The condition (\ref{condition}) determines the value of $d({\kappa})$ and then the 
dimension of ${\cal F}_{\kappa}$. The finiteness or infiniteness of ${\cal F}_{\kappa}$ depends on the 
sign of the parameter $\kappa$. For $\kappa \geq 0$, the space ${\cal F}_{\kappa}$ is infinite-dimensional. In 
fact, for $\kappa = 0$, the space ${\cal F}_{0}$ coincides with the usual Hilbert-Foch space for the harmonic oscillator. For 
$\kappa < 0$,  there exists a finite number of states satisfying the condition (\ref{condition}). As a matter of fact, for 
$\kappa < 0$, $n$ can take the values 
\begin{eqnarray}
n = 0, 1, \ldots, E(-\frac{1}{\kappa}) \equiv d-1, 
      \end{eqnarray}
where $E(x)$ stands for the integer part of $x$. The finiteness of the space ${\cal F}_{\kappa}$ induces properties 
of the operators $a^-$ and $a^+$ which differ from those corresponding to an infinite-dimensional space. In particular, 
the trace of any commutator in the finite-dimensional space must be zero. This implies that the parameter $\kappa$ is 
related to the dimension $d$ of the space ${\cal F}_{\kappa}$ by 
	  \begin{eqnarray} 
d = 1 - \frac{1}{\kappa}.
    \label{dimension} 
	  \end{eqnarray} 
Equation (\ref{dimension}) requires that $-1/\kappa$ be a positive integer. In the following, we shall assume that 
$-1/\kappa \in \mathbb{N}^*$ when $\kappa < 0$. 

 \subsection{A generalized oscillator Hamiltonian}
We are now in a position to define an operator which generalizes (up to an additive constant) the 
Hamiltonian $a^+ a^- + 1/2$ for the one-dimensional harmonic oscillator. Starting from 
      \begin{eqnarray}
a^+ a^- \vert n \rangle = F(n) \vert n \rangle \Rightarrow F(N) = a^+ a^-, 
      \end{eqnarray}
we refer $F(N)$ to as an Hamiltonian associated with the generalized oscillator algebra ${\cal A}_{\kappa}$. The
eigenvalue equation 
      \begin{eqnarray}
F(N) \vert n \rangle = n [1 + \kappa (n-1)] \vert n \rangle
      \end{eqnarray} 
gives the energies (\ref{eigenvalue}) of a quantum dynamical system described by the Hamiltonian operator $F(N)$. Let us 
discuss the degeneracies of the levels $F(n)$ given by (\ref{eigenvalue}). 

(i) In the case $\kappa \geq 0$, the spectrum of $F(N)$ is nondegenerate. 

(ii) In the case $\kappa < 0$, the eigenvalues of $F(N)$ can be rewritten as 
      \begin{eqnarray}
F(n) = n \frac{d-n}{d-1},
\label{eigenvalue avec d} 
      \end{eqnarray}
so that
      \begin{eqnarray}
F(n) = F(d-n) \qquad n = 1, 2, \ldots, d-1.
      \end{eqnarray}
Thus, for $d$ even the levels are doublets except the fundamental level $n=0$ and the level 
$n = d/2$ which are nondegenerate. For $d$ odd the levels are two-fold degenerate except the 
fundamental level $n=0$ which is a singlet. 

In both cases ($\kappa \geq 0$ and $\kappa < 0$), we note that the Perron-Frobenius theorem \cite{Reed} is satisfied, namely, 
the fundamental level is nondegenerate. 

It is known that one-dimensional quantum dynamical systems (on the real line) correspond to 
nondegenerate spectra. Therefore, the representation obtained for ${\cal A}_{\kappa}$ with 
$\kappa < 0$ cannot be used to describe a particle evolving in some nonrelativistic potential 
on the real line. However, a modification of the generalized oscillator algebra ${\cal A}_{\kappa}$
can be achieved in orded to avoid the degeneracies of $F(N)$. This will be done in section 4
by means of a truncation procedure which will prove also useful in the case $\kappa \geq 0$ 
to define in a consistent way the phase operator for some exactly solvable systems. 

\section{Temporally stable phase states for ${\cal A}_{\kappa}$}
We shall treat separately the cases $\kappa \geq 0$ and $\kappa < 0$ associated with the 
infinite- and the finite-dimensional representation of the generalized oscillator algebra 
${\cal A}_{\kappa}$, respectively. 

\subsection{The infinite-dimensional case}
In the case $\kappa \geq 0$, we decompose $a^-$ and $a^+$ as
      \begin{eqnarray} 
a^- = E_{\infty} \sqrt{F(N)} \qquad  a^+ = \sqrt{F(N)} \left( E_{\infty} \right)^{\dagger}, 
      \label{decompo cas infini}
      \end{eqnarray}
where
      \begin{eqnarray}
E_{\infty} := \sum_{n=0}^{\infty} e^{i [F(n+1)- F(n)] \varphi } \vert n \rangle \langle n+1 \vert.
      \end{eqnarray} 
It is important to emphasize that
      \begin{eqnarray}
E_{\infty}\left( E_{\infty} \right)^{\dagger} = \sum_{n=0}^{\infty} \vert n\rangle\langle n\vert = I \qquad 
\left( E_{\infty} \right)^{\dagger} E_{\infty}= \sum_{n=1}^{\infty} \vert n\rangle\langle n\vert = I - \vert 0 \rangle \langle 0 \vert, 
      \end{eqnarray} 
a result which means that $E_{\infty}$ is not a unitary operator. 

To find the phase states corresponding to $\kappa \geq 0$, 
let us consider the eigenvalue equation
            \begin{eqnarray}
E_{\infty} \vert z \rangle = z \vert z \rangle \qquad z \in \mathbb{C}.
            \end{eqnarray}
By expanding the vector $\vert z \rangle$ of ${\cal F}_{\kappa}$ as
      \begin{eqnarray}
\vert z \rangle = \sum_{n=0}^{\infty} C_n z^n \vert n \rangle,
      \end{eqnarray}
it is easy to see that the complex coefficients $C_n$ satisfy the relation
      \begin{eqnarray}
C_{n+1} = e^{-i [F(n+1) - F(n)] \varphi } C_n \qquad n \in \mathbb{N}.
      \end{eqnarray}
It follows that
      \begin{eqnarray}
C_n = e^{-i F(n) \varphi} C_0 \qquad n \in \mathbb{N}, 
      \end{eqnarray}
where the coefficient $C_0$ can be determined from the normalization condition of the states $\vert z \rangle$. As a result, 
we can take (up to a phase factor) 
      \begin{eqnarray}
\vert z \rangle = \sqrt{1 - |z|^2} \sum_{n=0}^{\infty} z^n e^{- i F(n) \varphi} \vert n \rangle
      \end{eqnarray}
on the domain $\{ z \in \mathbb{C}, |z| < 1 \}$. 

Following the method developed in \cite{VoudasBM} for the Lie algebra $su(1,1)$, we define the states $\vert \theta , \varphi \rangle$ by
       \begin{eqnarray}
 \vert \theta , \varphi \rangle := \lim_{z \rightarrow e^{i\theta}} \frac{1}{\sqrt{1 - |z|^2}} \vert z \rangle, 
       \end{eqnarray}
where $\theta \in [-\pi , +\pi]$ (see also \cite{VourdasLimit} where a limit of type 
$z \rightarrow e^{i\theta} \Rightarrow \vert z \vert \rightarrow 1$ is used in a similar way). We 
thus get the states
      \begin{eqnarray}
\vert \theta , \varphi \rangle = \sum_{n=0}^{\infty} e^{i n \theta} e^{- i F(n) \varphi} \vert n \rangle.
      \end{eqnarray}
These states, defined on the unit circle $S^1$, turn out to be phase states. Indeed, we have
      \begin{eqnarray}
E_{\infty} \vert \theta , \varphi \rangle  = e^{i \theta} \vert \theta , \varphi \rangle.
      \end{eqnarray}
Hence, the operator $E_{\infty}$ is a (nonunitary) phase operator.

The main properties of the states $\vert \theta , \varphi \rangle$ are the following.

(i) They are temporally stable in the sense that the relation
      \begin{eqnarray}
e^{-i F(N) t} \vert \theta , \varphi \rangle = \vert \theta , \varphi + t) 
      \end{eqnarray}
is satisfied for any value of the real parameter $t$. This property is due to 
the presence of the parameter $\varphi$ in the phase operator $E_{\infty}$.
 
(ii) They are not normalized and not orthogonal. However, for fixed $\varphi$, they satisfy the closure relation
      \begin{eqnarray}
\frac{1}{2\pi} \int_{-\pi}^{+\pi} d\theta \vert \theta , \varphi \rangle \langle \theta , \varphi \vert = I.
      \end{eqnarray}

Finally, observe that for $\varphi = 0$ the states $\vert \theta , 0 \rangle$ have the same form than those derived 
in \cite{VoudasBM} for $su(1,1)$. 

\subsection{The finite-dimensional case}
For $\kappa < 0$ with $-1/\kappa \in \mathbb{N}^*$, the Hilbert space 
${\cal F}_{\kappa}$ is $d$-dimensional with $d = 1 - 1/\kappa$. The action of $a^-$ and $a^+$ on 
${\cal F}_{\kappa}$ is given by (\ref{action sur les n}) supplemented by
      \begin{eqnarray}
a^+ \vert d-1 \rangle = 0, 
      \end{eqnarray}
which easily follows from the calculation of $\langle d-1 \vert a^- a^+ \vert d-1 \rangle$. 

Let us look for a decomposition of the creation $a^+$ and annihilation $a^-$ operators similar to 
(\ref{decompo cas infini}) for the case $\kappa \geq 0$. Thus, let us put
      \begin{eqnarray} 
a^- = E_{d} \sqrt{F(N)} \Leftrightarrow a^+ = \sqrt{F(N)} \left( E_{d} \right)^{\dagger}. 
      \label{decompo cas fini}
      \end{eqnarray}
The operator $E_{d}$ can be seen to satisfy
      \begin{eqnarray} 
E_d \vert n \rangle = e^{i [F(n) - F(n-1)] \varphi } \vert n-1 \rangle 
      \label{action de Ed}
      \end{eqnarray}
for $n = 1, 2, \ldots, d-1$. For $n=0$, we shall assume that 
            \begin{eqnarray} 
E_d \vert 0 \rangle = e^{i [F(0)- F(d-1)] \varphi} \vert d-1 \rangle 
            \end{eqnarray}  
so that (\ref{action de Ed}) is valid modulo $d$. (Note that, in view of (\ref{decompo cas fini}),  
$a^- \vert 0 \rangle = 0$ does not imply that $E_d \vert 0 \rangle = 0$.) It follows that we have
      \begin{eqnarray}
\left( E_d \right)^{\dagger} \vert n \rangle = e^{-i [F(n+1) - F(n)] \varphi } \vert n+1 \rangle, 
      \end{eqnarray}
where $n+1$ should be understood modulo $d$. As an important result (to be contrasted with the situtation where $\kappa \geq 0$), 
the operator $E_d$ is unitary. Therefore, equation (\ref{decompo cas fini}) constitutes a polar decomposition 
of $a^-$ and $a^+$.

We are now ready to derive the eigenstates of the operator $E_d$. Let us consider the eigenvalue equation
      \begin{eqnarray}
E_d \vert z \rangle = z \vert z \rangle \qquad \vert z \rangle = \sum_{n = 0}^{d-1} C_n z^n \vert n \rangle
      \end{eqnarray}
with $z \in \mathbb{C}$. Here again (as in the case $\kappa \geq 0$), we obtain a recurrence relation for 
the coefficients $C_n$, viz., 
            \begin{eqnarray}
C_{n} = e^{-i [F(n) - F(n-1)] \varphi} C_{n-1} \qquad n = 1, 2, \ldots, d-1
            \end{eqnarray}
with the cyclic condition
      \begin{eqnarray}
C_{0} = z^d e^{-i [F(0) - F(d-1)] \varphi} C_{d-1}.
      \end{eqnarray} 
Therefore, we get
      \begin{eqnarray}
C_n = e^{-i F(n) \varphi} C_0 \qquad n = 0, 1, \ldots, d-1,  
      \end{eqnarray}
with the discretization condition
      \begin{eqnarray} 
z^d = 1.
      \end{eqnarray}
As a consequence, the complex variable $z$ is a root of unity
given by
      \begin{eqnarray} 
z = q^m \qquad m = 0, 1, \ldots, d-1,
      \end{eqnarray}
where 
\begin{eqnarray}
q := e^{2 \pi i / d}
\label{definition of q}
\end{eqnarray}
is reminiscent of the parameter used in the theory of quantum groups. The constant $C_0$ 
can be calculated from the normalization condition $\langle z \vert z \rangle = 1$ to be 
      \begin{eqnarray}
C_0 = \frac{1}{\sqrt d}
      \end{eqnarray}
up to a phase factor. Finally, we arrive at the following 
eigenstates $\vert z \rangle \equiv \vert m , \varphi \rangle$ of $E_d$
\begin{eqnarray}
\vert m , \varphi \rangle = \frac{1}{\sqrt d} \sum_{n=0}^{d-1} e^{-i F(n) \varphi} q^{mn} \vert n \rangle. 
\label{coherentstatemvarphi}
\end{eqnarray}
The states $\vert m , \varphi \rangle$, labeled by the parameters $m \in \mathbb{Z}/d\mathbb{Z}$ and $\varphi \in \mathbb{R}$, satisfy 
      \begin{eqnarray}
E_d \vert m , \varphi \rangle = e^{i\theta_m} \vert m , \varphi \rangle \qquad \theta_m = m \frac{2 \pi}{d},
      \end{eqnarray}
which shows that $E_d$ is indeed a phase operator. In the particular case $\varphi = 0$, the states $\vert m , 0 \rangle$
are similar to those derived in \cite{VoudasBM} for the Lie algebra $su(2)$. In this case, the states $\vert m , 0 \rangle$ 
correspond to an ordinary discrete Fourier transform of the basis $\{ \vert n \rangle : n = 0, 1, \ldots, d-1 \}$ of the 
$d$-dimensional space ${\cal F}_{\kappa}$.

The phase states $\vert m , \varphi \rangle$ have remarkable properties (to be 
compared to those for the states $\vert \theta , \varphi \rangle$ of the case $\kappa \geq 0$). 

(i) They are temporally stable under ``time evolution''. In other words, they satisfy
      \begin{eqnarray}
e^{-i F(N) t} \vert m , \varphi \rangle = \vert m , \varphi + t \rangle. 
      \end{eqnarray} 
for any value of the real parameter $t$. We note here the major role of the parameter 
$\varphi$ in ensuing the temporal stability of the states $\vert m , \varphi \rangle$. 

(ii) For fixed $\varphi$, they satisfy the equiprobability relation
\begin{eqnarray}
| \langle n \vert m , \varphi \rangle | = \frac{1}{\sqrt{d}} \qquad n, m \in \mathbb{Z}/d\mathbb{Z}.
\label{computational et MUB}
\end{eqnarray}

(iii) For fixed $\varphi$, they satisfy the orthonormality relation
      \begin{eqnarray}
\langle m , \varphi \vert m' , \varphi \rangle = \delta_{m,m'} \qquad m, m' \in \mathbb{Z}/d\mathbb{Z}
      \end{eqnarray}
and the closure property
      \begin{eqnarray}
\sum_{m = 0}^{d-1} \vert m , \varphi \rangle \langle m , \varphi \vert = I.
      \end{eqnarray}

(iv) The overlap between two phase states $\vert m' , \varphi' \rangle$ and $\vert m , \varphi \rangle$ reads
   \begin{eqnarray}
\langle m , \varphi \vert m' , \varphi' ) = \frac{1}{d} \sum_{n=0}^{d-1} q^{\rho(m-m', \varphi - \varphi', n)}, 
   \label{overlap}
   \end{eqnarray}
where
         \begin{eqnarray}
\rho(m-m', \varphi - \varphi', n) = - (m - m')n + \frac{d}{2\pi} (\varphi - \varphi') F(n)
         \end{eqnarray}
and $q$ is defined in (\ref{definition of q}). Therefore, the temporally stable phase states are not all 
orthogonal.

\section{{Truncated generalized oscillator algebra and phase states}}

As discussed in section 2, in the case $\kappa \geq 0$ the Hilbert space ${\cal F}_{\kappa}$ associated with ${\cal A}_{\kappa}$ 
is infinite-dimensional. It is then impossible to define a unitary phase operator (see section 3). On the other hand, in the case $\kappa < 0$ 
with $- 1/ \kappa \in \mathbb{N}^*$ the space ${\cal F}_{\kappa}$ is finite-dimensional and there is no problem to define a 
unitary phase operator. However, the spectrum of the Hamiltonian $F(N)$ associated with ${\cal A}_{\kappa}$ for 
$- 1/ \kappa \in \mathbb{N}^*$ exhibits degeneracies. Therefore, it is appropriate to truncate the space ${\cal F}_{\kappa}$ 
for both $\kappa \geq 0$ and $\kappa < 0$ in order to get a subspace ${\cal F}_{\kappa, s}$ of dimension $s$ with the basis 
$\{ |n\rangle : n = 0, 1, \ldots, s-1\}$. For $\kappa \geq 0$, the truncation is done at $s$ sufficiently large (note that the 
difference $F(n+1) - F(n)$ between two consecutive states increases with $n$ for $\kappa > 0$ so that we can ignore, in a perturbative 
scheme, the states with $n$ large). For $\kappa < 0$, the truncation can be done at $s = (d+2)/2$ for $d$ even and at $s = (d+1)/2$ for 
$d$ odd (with $d$ given by (\ref{dimension})) in order to avoid the degeneracies of $F(N)$.
 
\subsection{{The truncated algebra ${\cal A}_{\kappa, s}$}}

Inspired by the work of Pegg and Barnett \cite{Pegg}, we
define the truncated generalized oscillator algebra ${\cal A}_{\kappa, s}$ through the three linear operators $b^-$, $b^+$ and $N$ 
satisfying the following relations
      \begin{eqnarray}
[b^- , b^+] = I + 2 \kappa N - F(s) \vert s-1 \rangle \langle s-1 \vert \quad 
[N , b^{\pm}] = \pm b^{\pm} \quad 
\left( b^- \right)^{\dagger} = b^+ \quad
N^{\dagger} = N. 
      \label{thetruncatedalgebra}
      \end{eqnarray}
The algebra ${\cal A}_{\kappa, s}$ generalizes the one introduced by Pegg and Barnett for the harmonic oscillator in their discussion 
of the phase operator for the single modes of the electromagnetic field \cite{Pegg}. Indeed, the algebra 
${\cal A}_{0, s}$, for $\kappa = 0$, is identical to the truncated oscillator algebra considered in \cite{Pegg}. 

Following the same approach as in subsection 2.2, we define a $s$-dimensional representation of ${\cal A}_{\kappa, s}$ (whatever 
the sign of $\kappa$ is) via the actions 
	  \begin{eqnarray} 
& & b^+ \vert n \rangle = \sqrt{F(n+1)} e^{{-i [F(n+1)- F(n)]  \varphi }} \vert n+1 \rangle, \nonumber \\ 
& & b^- \vert n \rangle = \sqrt{F(n)}   e^{{ i [F(n) - F(n-1)] \varphi }} \vert n-1 \rangle, \label{action des b sur les n} \\ 
& & b^- \vert 0 \rangle = 0 \qquad b^+ \vert s-1 \rangle = 0 \qquad
    N   \vert n \rangle = n \vert n \rangle
    \nonumber 
	  \end{eqnarray}   
for $n = 0, 1, \ldots, s-1$. Note that a further condition is necessary here, namely, the upper limit condition 
$b^+ \vert s-1 \rangle = 0$. It can be checked that the recurrence relation (\ref{recurrence}) is equally valid 
for ${\cal A}_{\kappa, s}$. Therefore, equations (\ref{eigenvalue}) and (\ref{eigenvalue avec d}) can be applied 
with $n = 0, 1, \ldots, s-1$. 

It is interesting to note that the creation and annihilation operators $b^-$ and $b^+$ 
satisfy (in the representation under consideration) the nilpotency relations
      \begin{eqnarray}
(b^-)^s = (b^+)^s = 0, 
      \end{eqnarray}
which are similar (for $s = k \in \mathbb{N} \setminus {0,1})$ to those 
describing the so-called $k$-fermions that are objects interpolating between 
fermions (for $k = 2$) and bosons (for $k \to \infty$) \cite{DaoHasKib}. 

\subsection{{Phase states for ${\cal A}_{\kappa, s}$}}

For the truncated algebra ${\cal A}_{\kappa, s}$ (corresponding to $d(\kappa)$ 
finite or infinite), the analog of the phase operator $E_d$ is the unitary operator
      \begin{eqnarray}
E_s := e^{i [F(0) - F(s-1)] \varphi } \vert s-1 \rangle \langle 0 \vert + 
                       \sum_{n=1}^{s-1} e^{i [F(n) - F(n-1)] \varphi } \vert n-1 \rangle \langle n \vert. 
      \end{eqnarray}
By using the same reasoning as in subsection 3.2, we obtain
      \begin{eqnarray}
E_s \vert m , \varphi \rangle = e^{i\theta_m} \vert m , \varphi \rangle \qquad \theta_m = m \frac{2 \pi}{s},
      \end{eqnarray}
where
	\begin{eqnarray}
\vert m , \varphi \rangle = \frac{1}{\sqrt s} \sum_{n=0}^{s-1} e^{-i F(n) \varphi} (q_s)^{mn} \vert n \rangle,
	\label{coherentstatemvarphi pour Akappas}
	\end{eqnarray}     
with $m \in \mathbb{Z} / s \mathbb{Z}$, $\varphi \in \mathbb{R}$ and $q_s$ given by 
            \begin{eqnarray}
q_s : = e^{2 \pi i / s}.
            \label{definition de qs}
            \end{eqnarray}
We are thus left with phase states $\vert m , \varphi \rangle$ associated with the phase operator $E_s$. These states 
satisfy the same properties as those for $E_d$ (see section 3.2) except that $d$ is replaced by $s$ in some places. 

\subsection{{A new type of discrete phase states}}

It is well known that, for quantum systems with a finite spectrum (like the Morse system) or for Lie algebras with 
finite-dimensional unitary representations (as for instance $su(2)$),  the construction of coherent states cannot be 
achieved by looking for the eigenstates of an annihilation operator \cite{GazeauK} or of a compact shift operator 
\cite{BarutGi}. 

For the algebra ${\cal A}_{\kappa, s}$ the difficulty inherent to the finiteness of the representation can be overcome 
as follows. We define the operator 
		\begin{eqnarray}
 V_s := b^-  +  \frac{(b^+)^{s-1}}{E(s-1)}, 
 		\label{operatorVs}
		\end{eqnarray}
where the function $E$ is defined via 
		\begin{eqnarray}
E(0) := 1 \qquad E(n) := F(1) F(2) \ldots F(n) \qquad n = 1, 2, \ldots, s-1.
		\end{eqnarray}	
The operator $V_s$ is an idempotent operator of order $s$ since 
		\begin{eqnarray}
(V_s)^s = I.
 		\label{idempotencyVs}
		\end{eqnarray}
Let us consider the eigenvalue equation
            \begin{eqnarray}
V_s \vert z \rangle = z \vert z \rangle \qquad 
\vert z \rangle = \sum_{n=0}^{s-1} C_n z^n \vert n \rangle \qquad z \in \mathbb{C}.
            \end{eqnarray}
By using (\ref{idempotencyVs}), we obtain that $z$ is discretized as 
            \begin{eqnarray}
z = (q_s)^\mu \qquad \mu \in \mathbb{Z}/s\mathbb{Z}, 
            \end{eqnarray}
with $q_s$ defined by (\ref{definition de qs}). Then, it is a simple matter to calculate the coefficients $C_n$ and to normalize 
the $\mu$- and $\varphi$-dependent states $| z \rangle \equiv | \mu , \varphi \rangle$. 
This leads to 
			\begin{eqnarray}
\vert \mu , \varphi \rangle = C_0 \sum_{n=0}^{s-1} \frac{1}{\sqrt{E(n)}} (q_s)^{n \mu} e^{-i F(n) \varphi} \vert n \rangle, 
      \label{phase states for Akappas}
			\end{eqnarray}
where the normalization factor $C_0$ is such that (up to a phase factor) 
			\begin{eqnarray}
{\it C_0}^{-2} = \sum_{n=0}^{s-1} \frac{1}{E(n)}.
			\end{eqnarray}
The states $| \mu , \varphi \rangle$ are temporally stable and are similar to the coherent 
states introduced by Gazeau and Klauder \cite{GazeauK} except that their labeling includes an integer and they 
correspond to the eigenvectors of a polynomial in terms of generalized creation and annhilation operators. They satisfy 
			\begin{eqnarray}
\langle \mu , \varphi \vert \mu' , \varphi' \rangle = C_0^2 \sum_{n=0}^{s-1} \frac{1}{E(n)} (q_s)^{n(\mu'-\mu)} e^{-i F(n) (\varphi' - \varphi)} 
			\end{eqnarray}
and 
			\begin{eqnarray}
\frac{1}{s} \sum_{m=0}^{s-1} \vert \mu , \varphi \rangle \langle \mu , \varphi \vert = 
C_0^2 \sum_{n=0}^{s-1} \frac{1}{E(n)} \vert n \rangle \langle n \vert.
			\end{eqnarray}
Consequently, they are not orthogonal.

We close this subsection with a remark concerning the unitary operator
			\begin{eqnarray}
U_s := (q_s)^N 
			\end{eqnarray}
that is a companion of $V_s$ in the following sense. This operator satisfies the cyclicity relation	
		\begin{eqnarray}
(U_s)^s = I.
 		\label{idempotencyUs}
		\end{eqnarray}
Furthermore, we have the $s$-commutation relation 
		\begin{eqnarray}
V_sU_s - q_s U_sV_s = 0.
 		\label{scommutation}
		\end{eqnarray}
Equations (\ref{idempotencyVs}), (\ref{idempotencyUs}) and (\ref{scommutation}) are necessary conditions 
for the pair ($U_s, V_s$) be a pair of Weyl (see \cite{KibJPhysA0809}). However, this is not the case 
because $V_s$ is not unitary.   
              
\section{Application to mutually unbiased bases}
As an {\it a priori} unexpected connection, the approach in subsection 3.2 and 4.2 for the finite-dimensional 
cases (for ${\cal A}_{\kappa}$ and ${\cal A}_{\kappa, s}$) can be further developed for deriving MUBs. Let us 
recall that two orthonormal bases $\{ \vert a \alpha \rangle : \alpha = 0, 1, \ldots, d-1 \}$ 
and $\{ \vert b \beta \rangle : \beta = 0, 1, \ldots, d-1 \}$ in a $d$-dimensional Hilbert space 
(with an inner product $\langle \, \vert \, \rangle$) are said to be mutually unbiased iff 
      \begin{eqnarray}
\vert \langle a \alpha \vert b \beta \rangle \vert = \delta_{a,b} \delta_{\alpha,\beta} + \frac{1}{\sqrt{d}} (1 - \delta_{a,b}).
      \end{eqnarray}
For fixed $d$, it is known that the number ${\cal N}$ of MUBs is such that ${\cal N} \leq d+1$ and that the limit ${\cal N}=d+1$ 
is reached when $d$ is the power of a prime number \cite{Ivanovic, WoottersFields}.

\subsection{MUBs from phase states for ${\cal A}_{\kappa}$}

In order to generate MUBs along the line of the developments of subsection 3.2, 
let us further examine some properties of the phase operator $E_d$ for ${\cal A}_{\kappa}$ 
with $\kappa < 0$. This operator can be written in a compact form as 
      \begin{eqnarray}
E_d = \sum_{n=0}^{d-1} e^{i [F(n) - F(n-1)] \varphi } \vert n-1 \rangle \langle n \vert
      \end{eqnarray}
(in this section, the summations on $n$ are understood modulo $d$). It is easy to check that 
      \begin{eqnarray}
\left( E_d \right)^d = I,
      \end{eqnarray}
so that $E_d$ is idempotent. The operator $E_d$ can be decomposed as 
      \begin{eqnarray} 
E_d = U_{\varphi} V,
      \end{eqnarray}
where the operators $U_{\varphi}$ and $V$ are defined by
      \begin{eqnarray} 
U_{\varphi} := e^{i [F(N+1) - F(N)] \varphi} \qquad V := \sum_{n=0}^{d-1} \vert n-1 \rangle \langle n \vert.
      \end{eqnarray}
The operators $U_{\varphi}$ and $V$ are unitary and satisfy the pseudo-commutation relation
\begin{eqnarray}
U_{\varphi} V = e^{2 i \varphi / (d-1)} V U_{\varphi}.
\label{pseudocommutation}
\end{eqnarray}
In addition, the operator $V$ satisfies the idempotency relation
      \begin{eqnarray}
V^d = I
      \end{eqnarray}
and, when the parameter $\varphi$ is quantized as 
\begin{eqnarray}
\varphi = - \pi \frac{d-1}{d} p \qquad p \in \mathbb{Z}/d\mathbb{Z},  
\label{phidiscrete}
\end{eqnarray}
we have
      \begin{eqnarray}
\left( U_{\varphi} \right)^d = e^{i \pi (d-1)p} I. 
      \end{eqnarray}
In view of (\ref{phidiscrete}), equation (\ref{pseudocommutation}) can be rewritten as 
      \begin{eqnarray}
VU_{\varphi} = q^p U_{\varphi}V
      \end{eqnarray}
(see (\ref{definition of q}) for the definition of $q$). For the discrete values of $\varphi$ afforded 
by (\ref{phidiscrete}), equation (\ref{coherentstatemvarphi}) yields the phase states 
$\vert m , \varphi \rangle \equiv \vert m , p \rangle$ given by  
	\begin{eqnarray}
\vert m , p \rangle = \frac{1}{\sqrt{d}} \sum_{n = 0}^{d-1} q^{n(d-n)p/2 + nm} \vert n \rangle 
\qquad p, m \in \mathbb{Z}/d\mathbb{Z}, 
	\label{MUBpm} 
	\end{eqnarray}
which coincides with the vector $\vert a \alpha \rangle$, with $a \equiv p$ and $\alpha \equiv m$, 
obtained in \cite{KibKPAlbKib} in an $SU(2)$ approach to MUBs. Alternatively, by putting 
      \begin{eqnarray}
k := d - n - 1 \qquad \vert n \rangle = \vert d - k - 1 \rangle \equiv \vert k \rangle, 
      \end{eqnarray}
equation (\ref{MUBpm}) becomes 
      \begin{eqnarray}
\vert m , p \rangle = \frac{1}{\sqrt{d}} \sum_{k = 0}^{d-1} q^{(k+1)(d-k-1)p/2 - (k+1)m} \vert k \rangle, 
\qquad p, m \in \mathbb{Z}/d\mathbb{Z},
      \label{soixante dix neuf}
      \end{eqnarray} 
which coincides with the vector $\vert a \alpha \rangle$, with $a \equiv p$ and $\alpha \equiv m$, 
derived in \cite{KibJPhysA0809} in an angular momentum approach to MUBs. It is to be observed that 
(\ref{MUBpm}) and (\ref{soixante dix neuf}) correspond to quadratic discrete Fourier transforms.  

To make a further contact with \cite{KibKPAlbKib, KibJPhysA0809}, let us note that when $\varphi$ 
is discretized according to (\ref{phidiscrete}), the inner product 
$\langle m , \varphi \vert m' , \varphi' \rangle \equiv \langle m , p \vert m' , p' \rangle$ 
(see equation (\ref{overlap})) can be rewritten as 
      \begin{eqnarray}
\langle m , p \vert m' , p' \rangle = 
\frac{1}{d} S(u, v, w)
\label{overlapMUB}
      \end{eqnarray}
with 
      \begin{eqnarray}
u := p-p' \qquad v := - (p-p')d + 2(m' - m) \qquad w := d. 
      \end{eqnarray}
In equation (\ref{overlapMUB}), the factor $S(u, v, w)$ denotes a generalized quadratic Gauss sum defined 
by \cite{les2Berndt}
      \begin{eqnarray}
S(u, v, w) := \sum_{k=0}^{|w|-1} e^{i \pi (uk^2 + vk) / w},
      \label{Gauss sum}
      \end{eqnarray}
where $u$, $v$ and $w$ are integers (the nonvanishing of $S(u,v,w)$ requires $uw + v$ even). In the special case where $d$ is a prime integer and $p' \not= p$, the calculation 
of $S(u, v, w)$ in (\ref{overlapMUB}) through the methods developed in \cite{les2Berndt, lesquatre} (see also \cite{KibKPAlbKib}) leads to 
      \begin{eqnarray}
\vert \langle m , p \vert m' , p' \rangle \vert = \frac{1}{\sqrt{d}}. 
      \end{eqnarray}
This result shows that the $d$ bases 
      \begin{eqnarray}
B_p := \{ | m , p \rangle : m = 0, 1, \ldots, d-1 \} \qquad p = 0, 1, \ldots, d-1
      \end{eqnarray}
of the $d$-dimensional space ${\cal F}_{\kappa}$, with $d$ given by (\ref{dimension}), are mutualy unbiased. On 
the other hand, in view of (\ref{computational et MUB}), it is clear that any basis $B_p$ and the basis 
      \begin{eqnarray}
B_d := \{ | n \rangle : n = 0, 1, \ldots, d-1 \}, 
      \end{eqnarray} 
known as the computational basis in quantum information and quantum computation, are mutually unbiased. As a conclusion, for 
$d$ prime, the $d$ bases $B_p$ with $p = 0, 1, \ldots, d-1$ and the computational basis $B_d$ constitute a complete set of 
$d+1$ MUBs. This result, in agreement with the one derived in \cite{KibKPAlbKib, KibJPhysA0809}, is the starting point for 
constructing MUBs in power prime dimension.

\subsection{MUBs from phase states for ${\cal A}_{\kappa, s}$}

By applying a discretization procedure similar to the one introduced in subsection 5.1, we 
can construct MUBs from the phase states (\ref{coherentstatemvarphi pour Akappas}) for 
the truncated algebra ${\cal A}_{\kappa , s}$ with $\kappa \not= 0$. Let us quantize the 
parameter $\varphi$ by putting 
      \begin{eqnarray}
\varphi = \pi \frac{2}{s \kappa} p \qquad p \in \mathbb{Z} / s \mathbb{Z}.
      \end{eqnarray}
Then, equation (\ref{coherentstatemvarphi pour Akappas}) leads to the states $| m , \varphi \rangle \equiv | m , p \rangle$ 
given by 
	\begin{eqnarray}
\vert m , p \rangle = \frac{1}{\sqrt{s}} \sum_{n = 0}^{s-1} (q_s)^{n(\delta - n)p + nm} \vert n \rangle 
\qquad p, m \in \mathbb{Z}/s\mathbb{Z}, 
	\label{MUBmp pour Akappas}
	\end{eqnarray}
where 
	\begin{eqnarray}
\delta := 1 - \frac{1}{\kappa}. 
	\end{eqnarray}
In this subsection, we shall assume that $1 / \kappa \in \mathbb{Z}$ so that $\delta \in \mathbb{Z}$ (note that $\delta = d$ for $\kappa < 0$). 
The overlap $\langle m , p \vert m' , p' \rangle$ can be written in terms of the generalized quadratic Gauss sum (\ref{Gauss sum}). Indeed, we have 
 		\begin{eqnarray}
\langle m , p \vert m' , p' \rangle = \frac{1}{s} \sum_{n=0}^{s-1} (q_s)^{n(\delta - n) (p' - p) + n(m' - m)} = \frac{1}{s} S(u, v, w), 
 		\end{eqnarray}
where
	\begin{eqnarray}
u := 2(p-p') \qquad v := 2 \delta (p'-p) + 2(m'-m) \qquad w := s.
	\end{eqnarray}
We can proceed as in subsection 5.1 in order to show that the various states $\vert m , p \rangle$ generate, together 
with the $s$-dimensional basis $\{ | n \rangle : n = 0, 1, \ldots, s-1 \}$, $s+1$ MUBs when $s$ is a prime integer. 

\section{Application to exactly solvable potentials}

The main goal of this section is to show how the generalized oscillator algebra ${\cal A}_{\kappa}$ is relevant
for the study of one-dimensional exactly solvable potentials in the context of supersymmetric quantum mechanics 
and how MUBs can be derived from the temporally stable phase states for some quantum mechanical systems.

\subsection{Creation, annihilation and transfer operators}

Ordinary supersymmetric quantum mechanics can be presented in differnt ways 
(e.g., see \cite{29W81}-\cite{MD26fresh}). We 
adopt here the approach according to which a supersymmetric dynamical system is defined by a triplet $(H, Q_+, Q_-)_2$ of 
linear operators acting on a $\mathbb{Z}_2$-graded Hilbert space ${\cal H}$ and satisfying the following 
relations
 	\begin{eqnarray}
 H = H^{\dagger} \qquad Q_- = Q_+^{\dagger} \qquad Q_{\pm}^2=0 \qquad
\{ Q_- , Q_+ \} =  H \qquad [ H , Q_{\pm}] = 0.
 	\label{relations triplet}
 	\end{eqnarray}
(In this approach, ordinary supersymmetric quantum mechanics is a particular case, corresponding to $k=2$, of fractional 
supersymmetric quantum mechanics of order $k$ dealing with triplets $(H, Q_+, Q_-)_k$ which satisfy relations 
generalizing (\ref{relations triplet}) and which correspond to a $\mathbb{Z}_k$ grading \cite{DaoKibPLAJMP}.) The 
operators $Q_+$ and $Q_-$ are the supercharges of the one-dimensional system. We suppose that the spectrum of the 
self-adjoint operator $H$, the supersymmetric Hamiltonian of the system, is discrete. The 
Hamiltonian $H$ can be written
 	\begin{eqnarray}
 	H = H_0 + H_1,
 	\end{eqnarray}
where $H_0$ and $H_1$ act on the states $\vert \Psi_n , 0 \rangle$ and $\vert \Psi_n , 1 \rangle$ of even and odd grading,
respectively. In other words, the Hilbert space ${\cal H}$ is decomposed as
 	\begin{eqnarray}
 	{\cal H} = {\cal H}_0 \oplus {\cal H}_1 \qquad
 	{\cal H}_0 := \{\vert \Psi_n , 0 \rangle : n \ {\rm ranging} \} \qquad
 	{\cal H}_1 := \{\vert \Psi_n , 1 \rangle : n \ {\rm ranging} \},
 	\end{eqnarray}
which reflects the ${\mathbb{Z}}_2$ grading. We shall assume that there is no supersymmetry breaking. In this case, 
the Hamiltonians $H_0$ and $H_1$ are isospectral except that the ground state of $H_0$ has no supersymmetric 
partner in the spectrum of $H_1$.

By combining the above-mentioned considerations on supersymmetry with the Infeld and Hull factorization method 
\cite{27Infeld}, we can construct creation, annihilation and transfer operators for an exactly solvable 
Hamiltonian in one dimension \cite{29W81}-\cite{russes}. For this purpose, let us consider a one-dimensional quantum system embedded 
in a real potential $v_0 : x \mapsto v_0(x)$. The corresponding Hamiltonian is 
 	\begin{eqnarray}
  H_0 := - \frac{1}{2} \frac{d^2}{dx^2} + v_0.
  \end{eqnarray}
Let us suppose that the Hamiltonian $H_0$ is exactly solvable and admits the discrete spectrum 
 	\begin{eqnarray}
e_0 = 0 < e_1 < e_2 < \ldots < e_n < e_{n+1} < \ldots, 
 	\end{eqnarray}
with a finite or infinite number of levels. We know that the Hamiltonian $H_0$ of this system can be factorized as 
\cite{31CKS9501, 30J96, 27Infeld, MD27Mielnik, MD26fresh}
 	\begin{eqnarray}
H_0 = x^+ x^- \qquad 
x^+ := \frac{1}{\sqrt{2}} \left( -\frac{d}{dx} + w \right) \qquad 
x^- := \frac{1}{\sqrt{2}} \left(  \frac{d}{dx} + w \right).
  \end{eqnarray}
The superpotential $w : x \mapsto w(x)$ satisfies the Ricatti equation
\begin{eqnarray} 
v_0 = \frac{1}{2} \left( w^2 - \frac{dw}{dx} \right).
\end{eqnarray}
Since the ground state energy is assumed to be zero, it is easy to see that the potential 
$v_0$ and the superpotential $w$ can be expressed in terms of the ground state wavefunction.

It is important to stress that the operators $x^+$ and $x^-$ are not in general creation and annihilation operators for $H_0$ 
\cite{31CKS9501, 30J96, MD27Mielnik, MD26fresh, russes}. They 
are indeed transfer operators from the spectrum of $H_0$ to the one of $H_1$ and vice-versa. To identify them, we start by 
representing the supercharge operators and the supersymmetric Hamiltonian by $2 \times 2$ matrices 
\cite{31CKS9501, 30J96, 32CGK04, MD26fresh}
\begin{eqnarray}
Q_- = 
\pmatrix{
0 & x^- \cr
0 & 0   \cr
}
\qquad
Q_+ =
\pmatrix{
0   & 0 \cr
x^+ & 0 \cr
}
\qquad
H =
\pmatrix{
H_1 & 0   \cr
0   & H_0 \cr
}, 
\end{eqnarray}
where the operator
\begin{eqnarray}
H_1 = x^-x^+ = - \frac{1}{2} \frac{d^2}{dx^2} + v_1
\end{eqnarray}
is the supersymmetric partner of $H_0$ and corresponds to a new potential $v_1 : x \mapsto v_1(x)$. The potential
\begin{eqnarray}
v_1 = \frac{1}{2} \left( w^2 + \frac{dw}{dx} \right) 
\end{eqnarray}
is the supersymmetric partner of the potential $v_0$. The Hamiltonian $H_1$ is also exactly solvable and isospectral to $H_0$ (except for the ground state). Indeed, 
    \begin{eqnarray} 
H_0 \vert \Psi_n , 0 \rangle = e^0_n \vert \Psi_n , 0 \rangle \Rightarrow 
H_1 (x^- \vert \Psi_n , 0 \rangle) = e^0_n (x^- \vert \Psi_n , 0 \rangle),
    \label{action de H1}
    \end{eqnarray}
where $e^0_n := e_n$. Similarly, 
    \begin{eqnarray} 
H_1 \vert \Psi_n , 1 \rangle = e^1_n \vert \Psi_n , 1 \rangle \Rightarrow 
H_0 (x^+ \vert \Psi_n , 1 \rangle) = e^1_n (x^+ \vert \Psi_n , 1 \rangle).
    \label{action de H0}
    \end{eqnarray}
(For more details see \cite{31CKS9501, 30J96, MD27Mielnik} and the recent topical review \cite{MD26fresh}.)
From equations (\ref{action de H1}) and (\ref{action de H0}), it is clear that we can take
\begin{eqnarray}
x^- \vert \Psi_{n+1} , 0 \rangle = \sqrt{e_{n+1}}e^{i (e_{n+1}- e_{n}) \varphi} \vert \Psi_n , 1 \rangle
\label{action de xmoins}
\end{eqnarray}
\begin{eqnarray}
x^+ \vert \Psi_n , 1 \rangle     = \sqrt{e_{n+1}}e^{-i (e_{n+1}- e_{n}) \varphi} \vert \Psi_{n+1} , 0 \rangle,
\label{action de xplus}
\end{eqnarray}
where $\varphi$ is a real number, and that the energies of the supersymmetric partners $H_0$ and $H_1$ are related by
\begin{eqnarray}
e^1_n = e^0_{n+1}.
\end{eqnarray}
Note that the operator $x^-$ (respectively $x^+$) converts an eigenfunction of $H_0$ (respectively $H_1$) into an 
eigenfunction of $H_1$ (respectively $H_0$) with the same energy. Thus, the operators $x^-$ and $x^+$ transfer the 
states from one spectrum to its partner spectrum. To introduce the ladder operators inside a given spectrum, we 
first consider the unitary operator $U$ relating the states $\vert \Psi_n , 0 \rangle$ and $\vert \Psi_n , 1 \rangle$ 
through (cf \cite{MD28Faddev}-\cite{MD29Samsonov})
\begin{eqnarray} 
U := \sum_{n} \vert \Psi_n , 1 \rangle \langle \Psi_n , 0 \vert \Rightarrow \vert \Psi_n , 1 \rangle  = U \vert \Psi_n , 0 \rangle. 
\end{eqnarray}
Operators similar to $U$ were already considered for continuous spectra \cite{MD28Faddev, MD29Samsonov} and for discrete 
spectra \cite{MD30Pursey, MD31Kumar}. Then, we define the operators \cite{MD30Pursey}-\cite{MD31Kumar}
\begin{eqnarray}
a^+ := x^+ U \qquad a^- := U^{\dagger}x^-.
\end{eqnarray}
By using equations (\ref{action de xmoins}) et (\ref{action de xplus}), we obtain 
\begin{eqnarray}
a^- \vert \Psi_n , 0 \rangle = \sqrt{e_{n}}   e^{ i (e_{n}- e_{n-1}) \varphi } \vert \Psi_{n-1} , 0 \rangle
\label{action de amoins sur Psin0}
\end{eqnarray}
\begin{eqnarray}
a^+ \vert \Psi_n , 0 \rangle = \sqrt{e_{n+1}} e^{-i (e_{n+1}- e_{n}) \varphi } \vert \Psi_{n+1} , 0 \rangle.
\label{action de aplus sur Psin0}
\end{eqnarray}
Consequently, $a^+$ and $a^-$ are creation and annihilation operators for the Hamiltonian $H_0$. Furthermore, 
it is easily seen that
\begin{eqnarray}
a^+ a^- = x^+ x^- = H_0.
\label{a x H0}
\end{eqnarray}
Ladder operators for the Hamiltonian $H_1$ can be introduced in a similar way. 

\subsection{Physical realizations of the generalized oscillator algebra}

To simplify the notation, we set $\vert \Psi_n \rangle := \vert \Psi_n , 0 \rangle$.
From equations (\ref{action de amoins sur Psin0}) et (\ref{action de aplus sur Psin0}), we get 
 \begin{eqnarray}
 [ a^- , a^+] \vert \Psi_n \rangle = (e_{n+1} - e_n) \vert \Psi_n \rangle. 
 \end{eqnarray}
The number operator $N$ defined by
 \begin{eqnarray}
 N \vert \Psi_n \rangle = n \vert \Psi_n \rangle
 \end{eqnarray}
is in general (for an arbitrary quantum system) different from the product $a^+a^-$. 
Let us consider the situation where the creation 
and annihilation operators satisfy the commutation relation
\begin{eqnarray}
[ a^- , a^+] = a N + b,  
\end{eqnarray}
a relation used in the study of the so-called polynomial Heisenberg algebra introduced in \cite{MD32Fernandez}. In 
other words, we assume that the energy gap $e_{n+1} - e_n$ between two succussive levels is linear in $n$, i.e.
\begin{eqnarray}
e_{n+1} - e_n = a n + b, 
\end{eqnarray}
where $a$ and $b$ are two real parameters. We also assume that the eigenvalues of the operator $aN+b$ 
are positive. With these choices, the
algebra generated by the operators $a^+$, $a^-$ and $N$ is identical to the generalized oscillator algebra 
${\cal A}_{\kappa}$ modulo the replacements 
\begin{eqnarray} 
a^{\pm} \rightarrow \frac{a^{\pm}}{\sqrt{b}} \qquad \kappa \rightarrow \frac{1}{2} \frac{a}{b}
\end{eqnarray} 
in equation (\ref{thealgebra}). Thus, from equations (\ref{action de amoins sur Psin0}-\ref{a x H0}), we have
\begin{eqnarray}
H_0 = a^+ a^- = \frac{1}{2} a N(N-1) + b N.
\end{eqnarray}
For $a \not= 0$, the spectrum of $H_0$ is non-linear and is given by
 \begin{eqnarray}
 H_0 \vert \Psi_n \rangle =  e_n\vert \Psi_n \rangle \qquad e_n  =  \frac{1}{2} a n(n-1) + b n.
 \label{add124}
 \end{eqnarray}
Particular realizations of (\ref{add124}) in terms of one-dimensional solvable potentials were previously considered 
in \cite{Antoine, Kinani, DaoKibPLAJMP, MD34Quesne99, MD35Angelova}. Following the developments in \cite{DaoKibPLAJMP}, 
we consider the following remarkable cases.
 
(i) For ($a = 0$, $b > 0$), the spectrum of $H_0$
is infinite-dimensional ($n \in \mathbb{N}$) and does not present degeneracies. 

(ii) For ($a > 0$, $b \geq 0$), the spectrum of $H_0$
is infinite-dimensional ($n \in \mathbb{N}$) and does not present degeneracies.

(iii) For ($a < 0$, $b \geq 0$), the spectrum of $H_0$ is finite-dimensional with $n = 0, 1, \ldots, s-1$ where
\begin{eqnarray} 
s & = & -\frac{b}{a} + \frac{3}{2} \ {\rm for} \ - 2\frac{b}{a} \ {\rm odd} \\
s & = & -\frac{b}{a} + 1 \           {\rm for} \ - 2\frac{b}{a} \ {\rm even},  
\end{eqnarray}
and all the states are nondegenerate. 

It is possible to find a realization of each of the three cases above in terms of exactly solvable dynamical systems in one dimension. We give below the corresponding potential $v_0$ and transfer operators. 

(i) The case ($a = 0$, $b = 1$) corresponds to the harmonic oscillator 
(for which $n \in \mathbb{N}$) with 
\begin{eqnarray} 
v_0(x) = \frac{1}{2} (x^2 - 1)
\end{eqnarray}
and 
\begin{eqnarray}
x^{\pm} \equiv a^{\pm} = \frac{1}{\sqrt{2}} \left( \mp \frac{d}{dx} + x \right).
\end{eqnarray}
(For the harmonic oscillator, $U$ reduces to the identity operator.) 

(ii) The case ($a = 1$, $2b = u+v+1$), with $u > 1$ and $v >1 $, corresponds to the P\" oschl-Teller potential 
(for which $n \in \mathbb{N}$) with 
\begin{eqnarray}
v_0(x) = \frac{1}{8} \bigg[\frac{u(u-1)}{\sin^2 \frac{x}{2}} + \frac{v(v-1)}{\cos^2 \frac{x}{2}}\bigg] -\frac{1}{8}(u+v)^2 
\end{eqnarray}
and
\begin{eqnarray}
x^{\pm} = \frac{1}{\sqrt{2}} \left[ \mp \frac{d}{dx} + \frac{1}{2}\bigg(u \cot\frac{x}{2} - v \tan\frac{x}{2}\bigg) \right].
\end{eqnarray}

(iii) The case ($a = -1$, $2b = 2l-1$), with $l \in \mathbb{N}^*$, corresponds to the Morse potential 
(for which $n = 0, 1, \ldots, l$) with 
\begin{eqnarray}
v_0(x) =  \frac{1}{2} \left[ e^{-2x} - (2l+1)e^{-x} + l^2 \right],
\end{eqnarray}
and
\begin{eqnarray}
x^{\pm} = \frac{1}{\sqrt{2}} \left( \mp \frac{d}{dx} + l - e^{-x} \right).
\end{eqnarray}

\subsection{{Phase states and MUB for exactly solvable systems}}

From equation (\ref{coherentstatemvarphi pour Akappas}), we can obtain the phase states for a general 
quantum system described by a truncated generalized oscillator algebra ${\cal A}_{\kappa , s}$. We get 
  \begin{eqnarray}
\vert m , \varphi \rangle = \frac{1}{\sqrt s} \sum_{n=0}^{s-1} e^{-i e_n \varphi} (q_s)^{nm} \vert \Psi_n \rangle, 
	\label{coherentstatemvarphi pour Akappasbis}
	\end{eqnarray}      
with $s$ sufficiently large for the harmonic oscillator and the 
P\" oschl-Teller systems and $s = l+1$ for the Morse system. Furthermore, 
equation (\ref{MUBmp pour Akappas}) provides with a mean to generate MUBs associated with 
the cases (i), (ii) and (iii) of subsection 6.2. 

On the other hand, the discrete phase state (\ref{phase states for Akappas}) reads here 
			\begin{eqnarray}
\vert \mu , \varphi \rangle = C_0 \sum_{n=0}^{s-1} \frac{1}{\sqrt{E(n)}} e^{-i e_n \varphi} (q_s)^{n \mu} \vert \Psi_n \rangle, 
      \label{phase states for Akappas avec exactly s s}
			\end{eqnarray}
where the factor $E(n)$ can be calculated in the different cases (i), (ii) and (iii). A simple calculation
gives the following results in term of the $\Gamma$ function.

(i) For the harmonic oscillator potential:
\begin{eqnarray}
E(n) = \Gamma(n+1).
\end{eqnarray}

(ii) For the P\" oschl-Teller potential:
\begin{eqnarray} 
E(n)= \frac{\Gamma(n+1)\Gamma(n+u+v+1)}{2^n \Gamma(u+v+1)}.
\end{eqnarray}

(iii) For the Morse potential: 
\begin{eqnarray}
E(n) = \frac{\Gamma(n+1)\Gamma(2l)}{2^n \Gamma(2l-n)}.
\end{eqnarray}

It should be mentioned that the discrete phase states given by (\ref{phase states for Akappas avec exactly s s}) 
differ from the coherent states for exactly sovable potentials derived in \cite{Kinani, MD31Kumar, MD32Fernandez, MD35Angelova, MD36Bagrov} 
from supersymmetric quantum mechanics techniques. The noticeable difference comes from the fact that the states 
(\ref{phase states for Akappas avec exactly s s}) are temporally stable and are labeled by an integer instead of  
a continuous complex variable as in the coherent states derived in \cite{Kinani, MD31Kumar, MD32Fernandez, MD35Angelova, MD36Bagrov}. 
The states (\ref{phase states for Akappas avec exactly s s}) are eigenstates of the operator (\ref{operatorVs}) whereas the coherent states 
in \cite{Kinani, MD31Kumar, MD32Fernandez, MD35Angelova, MD36Bagrov} are obtained from the three standard definitions (involving 
annihilation operator, displacement operator, and uncertainty relation). 

\section{Concluding remarks}

The starting point of this article is based on the definition of a generalized oscillator algebra ${\cal A}_{\kappa}$. This
algebra is interesting in two respects. First, it describes in an unified way some exactly solvable one-dimensional 
systems having a nonlinear spectrum (for $\kappa \not= 0$) or a linear spectrum (for $\kappa = 0$). As typical examples, these quantum systems correspond to the P\" oschl-Teller potential (for $\kappa > 0$), the Morse potential (for $\kappa 
< 0$) and the infinite square well potential (for $\kappa = 1/3$) in addition to the harmonic oscillator potential (for 
$\kappa = 0$). Second, the algebra ${\cal A}_{\kappa}$ can take into account some nonlinear effects that may occur in 
the quantum description of quantized modes of the electromagnetic field (cf. \cite{Walls}).
 
In connection with the algebra ${\cal A}_{\kappa}$, the present work adresses three problems: the construction of a 
phase operator, the determination of its temporally stable eigenstates (the so-called phase states) and the derivation 
of MUBs from the obtained phase states. This is the first time that a connection between MUBs and dynamical systems is 
established. In this regard, the character "temporally stable" of the eigenstates of the phase operator is essential for the derivation of MUBs. The main results of this paper are as follows. 

For the case $\kappa \geq 0$ (which corresponds to an infinite representation of ${\cal A}_{\kappa}$), the phase operator is not unitary. We note in passing that the corresponding phase states are similar to those derived in \cite{VoudasBM} except that our states are temporally stable. However for 
$\kappa \geq 0$, by making a ({\it \`a la} Pegg and Barnett) truncation, which gives rise to a truncated generalized 
oscillator algebra ${\cal A}_{\kappa , s}$, we can define a unitary phase operator whose eigenstates lead to MUBs. 

For the case $\kappa < 0$ (which corresponds to a finite representation of ${\cal A}_{\kappa}$), it is possible to 
construct a unitary phase operator whose eigenstates are temporally stable. MUBs can be derived as a subset of 
these states. For $\kappa < 0$, the consideration of a truncated generalized oscillator algebra ${\cal A}_{\kappa , s}$ 
is nevertheless necessary in order to establish a connection with the Morse system and to derive associated MUBs. 

As a conclusion, in both cases ($\kappa \geq 0$ and $\kappa < 0$), the truncation procedure makes it possible to 
define a unitary phase operator for exactly solvable systems and to generate temporally stable phase states from 
which MUBs can be derived.

Another result of this paper concerns a new type of phase states. These temporally stable phase states, namely 
the states (\ref{phase states for Akappas}), are associated with the truncated algebra ${\cal A}_{\kappa , s}$. 
They are eigenstates of an operator defined in the enveloping algebra of ${\cal A}_{\kappa , s}$ and constitute discrete 
analogs of the coherent states derived in \cite{GazeauK}. More generally, this result shows that it is possible, 
for a finite spectrum, to derive new phase states similar to the coherent states of \cite{GazeauK} constructed, 
for an infinite spectrum, as eigenstates of an annihilation operator. The key of the derivation of the new states 
(for a finite spectrum) is to add a power of the creation operator to the annihilation operator. 

To close this paper, let us mention that the concept of MUBs was recently extended to infinite-dimensional 
Hilbert spaces \cite{Weigert}. In this vein, it is hoped that the temporally stable phase states derived in this 
work for the infinite-dimensional case could serve as a hint for deriving MUBs for continuous variables, a 
difficult challenge. 

\section*{Acknowledgments}
One of the authors (M D) would like to thank the hospitality and kindness extended to him by the {\it Groupe de physique 
th\'eorique de l'Institut de Physique Nucl\'eaire de Lyon} where this work was done. The other author (M R K) is grateful 
to Michel Capdequi-Peyran\`ere for useful comments. Thanks are due to one Referee and to the Adjudicator for constructive 
suggestions.


\newpage

\baselineskip=14pt

\begin{thebibliography}{99}

\bibitem{Louisell63}
Louisell W H 1963 {\it Phys. Lett.} {\bf 7} 60

\bibitem{Susskind64} 
Susskind L and Glogower J 1964 {\it Physics (U S)} {\bf 1} 49

\bibitem{Susskind68} 
Carruthers P and Nieto M M 1968 {\it Rev. Mod. Phys.} {\bf 40} 411

\bibitem{Pegg} 
Pegg D T and Barnett S M 1989 {\it Phys. Rev.} A {\bf 39} 1665 

\bibitem{Vourdas90} 
Vourdas A 1990 {\it Phys. Rev.} A {\bf 41} 1653

\bibitem{BarutGi}
Barut A O and Girardello L 1971 {\it Commun. Math. Phys.} {\bf 21} 41 

\bibitem{Perelomov86} 
Perelomov A M 1986 {\it Generalized Coherent States and Their Applications} (Berlin: Springer)

\bibitem{GazeauK}
Gazeau J-P and Klauder J R 1999 {\it J. Phys. A: Math. Gen.} {\bf 32} 123    
  
\bibitem{Antoine} 
Antoine J-P, Gazeau J-P, Monceau P, Klauder J R and Penson K A 2001 {\it J. Math. Phys.} {\bf  42} 2349

\bibitem{Kinani} 
El Kinani A H and Daoud M 2001 {\it J. Phys. A: Math. Gen.} {\bf 34} 5373

Daoud M and Hussin V 2002 {\it J. Phys. A: Math. Gen.} {\bf 35} 7381

El Kinani A H and Daoud M 2002 {\it J. Math. Phys.} {\bf 43} 714

\bibitem{DaoKib12} 
Daoud M and Kibler M R 2002 A fractional supersymmetric oscillator and its coherent
states in: {\it Proceedings of the Sixth International Wigner Symposium 1999} 
(Istanbul: Bogazici University Press)

Daoud M and Kibler M R 2001 On fractional supersymmetric quantum mechanics: the
fractional supersymmetric oscillator, in: {\it Symmetry and Structural Properties of 
Condensed Matter} eds T Lulek, B Lulek and A Wal (Singapore: World Scientific)

\bibitem{DaoKibPLAJMP} 
Daoud M and Kibler M 2004 {\it Phys. Lett.} A {\bf 321} 147

Daoud M and Kibler M R 2006 {\it J. Math. Phys.} {\bf 47} 122108

\bibitem{KibKPAlbKib} 
Kibler M R 2006 
{\it Int. J. Mod. Phys.} B {\bf 20} 1792

Kibler M R and Planat M 2006 {\it Int. J. Mod. Phys.} B {\bf 20} 1802

Albouy O and Kibler M R 2007 {\it SIGMA} {\bf 3} article 076 

\bibitem{KibJPhysA0809} 
Kibler M R 2008 {\it J. Phys. A: Math. Theor.} {\bf 41} 375302 

Kibler M R 2009 {\it J. Phys. A: Math. Theor.} {\bf 42} 353001

\bibitem{MD33LesQuene} 
Quesne C and Vansteenkiste N 1998 {\it Phys. Lett.} A {\bf 240} 21 

Quesne C 2000 {\it Phys. Lett.} A {\bf 272} 313; erratum 2000 {\it Phys. Lett.} A {\bf 275} 313 

Quesne C and Vansteenkiste N 2000 {\it Int. J. Theor. Phys.} {\bf 39} 1175

Quesne C 2003 {\it Mod. Phys. Lett.} A {\bf 18} 515

\bibitem{Reed} 
Reed M and Simon B 1978 {\it Methods of modern mathematical physics, 
Analysis of operators} vol 4 (New York: Academic Press) 

\bibitem{VoudasBM} 
Vourdas A, Brif C and Mann A 1996  
{\it J. Phys. A: Math. Gen.} {\bf 29} 5887 

\bibitem{VourdasLimit} 
Vourdas A 1993 {\it Phys. Scr.} {\bf 48} 84 

\bibitem{DaoHasKib} 
Daoud M, Hassouni Y and Kibler M 1998 
The $k$-fermions as objects interpolating between fermions and bosons
{\it Symmetries in Science X}
eds B Gruber and M Ramek (New York: Plenum Press)

Daoud M, Hassouni Y and Kibler M 1998 
{\it Phys. Atom. Nuclei} {\bf 61} 1821 

\bibitem{Ivanovic} 
Ivanovi\'c I D 1981 {\it J. Phys. A: Math. Gen.} {\bf 14} 3241

\bibitem{WoottersFields} 
Wootters W K and Fields B D 1989 {\it Ann. Phys. (N Y)} {\bf 191} 363

\bibitem{les2Berndt} 
Berndt B C and Evans R J 1981
{\it Bull. Am. Math. Soc.} {\bf 5} 107 

Berndt B C, Evans R J and Williams K S 1998 
{\it Gauss and Jacobi Sums} (New York: Wiley)   

\bibitem{lesquatre} 
Hannay J H and Berry M V 1980 {\it Physica D} {\bf 1} 267

Matsutani S and \^Onishi Y 2003 {\it Found. Phys. Lett.} {\bf 16} 325

Rosu H C, Trevi\~no J P, Cabrera H and Murgu\'ia J S 2006  
{\it Int. J. Mod. Phys.} B {\bf 20} 1860

Merkel W, Crasser O, Haug F, Lutz E, Mack H, Freyberger M, Schleich W P, 
Averbukh I, Bienert M, Girard B, Maier H and Paulus G G 2006
{\it Int. J. Mod. Phys.} B {\bf 20} 1893

\bibitem{29W81}
Witten E 1981 {\it Nucl. Phys. B} {\bf 188} 513

\bibitem{31CKS9501}
Cooper F, Khare A and Sukhatme U 1995 {\it Phys. Rep.} {\bf 251} 267 

Cooper F, Khare A and Sukhatme U 2001 {\it Supersymmetry in Quantum Mechanics}
(Singapore: World Scientific)

\bibitem{30J96}
Junker G 1996 {\it Supersymmetric Methods in Quantum and Statistical Physics} (Berlin: Springer)

\bibitem{32CGK04}
Combescure M, Gieres F and Kibler M 2004 {\it J. Phys. A: Math. Gen.} {\bf 37} 10385

\bibitem{27Infeld} 
Infeld L and Hull T E 1951 {\it Rev. Mod. Phys.} {\bf 23} 21

\bibitem{MD27Mielnik} 
Mielnik B 1984 {\it J. Math. Phys.} {\bf 25} 3387

Negro J, Nieto L M and Rosas-Ortiz O 2000 {\it J. Phys. A: Math. Gen.} {\bf 33} 7207

Mielnik B and Rosas-Ortiz O 2004 {\it J. Phys. A: Math. Gen.} {\bf 37} 10007

\bibitem{MD26fresh} Fernandez C D J 2009 {\it Preprint} arXiv:0910.0192

\bibitem{MD28Faddev} 
Faddeev L D 1963 {\it J. Math. Phys.} {\bf 4} 72

\bibitem{MD30Pursey}
Pursey D L 1986 {\it Phys. Rev.} D {\bf 33} 2267

\bibitem{MD29Samsonov}
Samsonov B F 2000 {\it J. Phys. A: Math. Gen.} {\bf 33} 591

\bibitem{MD31Kumar} 
Kumar M S and Khare A 1996 {\it Phys. Lett.} A {\bf 217} 73 

\bibitem{russes} 
Bagrov V G and Samsonov B F 1997 {\it Phys. Part. Nucl.} {\bf 28} 374

\bibitem{MD32Fernandez} 
Fernandez C D J, Nieto L M and Rosas-Ortiz O 1995 {\it J. Phys. A: Math. Gen.} {\bf 28} 2693  

Fernandez C D J and Hussin V 1999 {\it J. Phys. A: Math. Gen.} {\bf 32} 3603 

Carballo J M, Fernandez C D J, Negro J and Nieto L M 2004 {\it J. Phys. A: Math. Gen.} {\bf 37} 10349  

\bibitem{MD34Quesne99} 
Quesne C 1999 {\it J. Phys. A: Math. Gen.} {\bf 32}  6705

\bibitem{MD35Angelova} 
Angelova M and  Hussin V 2008 {\it J. Phys. A: Math. Gen.} {\bf 41} 304016

\bibitem{MD36Bagrov}
Bagrov V G and Samsonov B F 1996 {\it J. Phys. A: Math. Gen.} {\bf 29} 1011

\bibitem{Walls}
Walls D F 1983 {\it Nature} {\bf 306} 141 

\bibitem{Weigert} 
Weigert S and Wilkinson M 2008 {\it Phys. Rev.} A {\bf 78} 020303


\end{thebibliography}
\end{document}